\documentclass[letterpaper, 10 pt, conference]{ieeeconf}
\IEEEoverridecommandlockouts 
\usepackage{arydshln}
\usepackage[T1]{fontenc}
\usepackage{caption,color}
\usepackage{graphicx}
\usepackage{amsmath,amssymb,amsfonts}
\usepackage{textcomp}
\usepackage{dsfont}
\usepackage{bbold}
\usepackage{arydshln}
\usepackage{comment}
\usepackage{mathtools} 
\usepackage{mathrsfs}
\usepackage{enumerate}
\everymath{\displaystyle}
\usepackage{cases}
\usepackage[framed,thmmarks]{ntheorem}  
\usepackage[overload]{empheq}
\usepackage{booktabs}
\usepackage{datetime}
\usepackage{cite}
\usepackage{stfloats}
\usepackage{pifont}
\newenvironment{proofa}{{\noindent\bfseries{Proof:}}}{\hfill$\blacksquare$}
\usepackage{etoolbox}

\newtheorem{lemma}{Lemma}
\newtheorem{theorem}{Theorem}
\newtheorem{remark}{Remark}
\newtheorem{definition}{Definition}
\AtEndEnvironment{problem}{\hfill$\Box$}
\AtEndEnvironment{lemma}{\hfill$\Box$}
\AtEndEnvironment{theorem}{\hfill$\Box$}
\AtEndEnvironment{assumption}{\hfill$\Box$}
\AtEndEnvironment{condition}{\hfill$\Box$}
\AtEndEnvironment{remark}{\hfill$\Box$}
\AtEndEnvironment{corollary}{\hfill$\Box$}
\AtEndEnvironment{stassumption}{\hfill$\Box$}
\AtEndEnvironment{definition}{\hfill$\Box$}

\newcommand{\RomanNumeralCaps}[1]
    {\MakeUppercase{\romannumeral #1}}

\title{\textbf{A General Nonlinear Observer Design for Inertial Navigation Systems with Almost Global Stability Guarantees}}

\def\namedlabel#1#2{\begingroup
    #2%
    \def\@currentlabel{#2}%
    \phantomsection\label{#1}\endgroup
}
\author{Sifeddine Benahmed, Tarek Hamel, Soulaimane Berkane
\thanks{Sifeddine Benahmed is with the Department of Technology \& Innovation, Capgemini Engineering, Toulouse, 31300, France (sif-eddine.benahmed@capgemini.com).}
\thanks{Soulaimane Berkane is with the Department of Computer Science and Engineering, University of Quebec in Outaouais, QC J8X 3X7, Canada (soulaimane.berkane@uqo.ca).}
\thanks{Tarek Hamel is with I3S-UniCA-CNRS, University Cote d’Azur, and the Institut Universitaire de France, France
(thamel@i3s.unice.fr)}
\thanks{* This research work is supported in part by NSERC-DG RGPIN-2020-04759 and the Fonds de recherche du Qu\'ebec (FRQ).}
\thanks{}}

\newcommand{\RNum}[1]{\uppercase\expandafter{\romannumeral #1\relax}}

\newcommand{\R}{\mathds{R}}

\newcommand{\Zsg}{\mathds{Z}_{>0}}

\newcommand{\sphere}{\mathds{S}}

\newcommand{\iframe}{\{\mathcal{I}\}}
\newcommand{\bframe}{\{\mathcal{B}\}}
\newcommand{\so}{\mathrm{SO}(3)}
\newcommand{\wx}{[\omega]_{\times}}
\newcommand{\Rthree}{\R^{3}}
\newcommand{\liealg}{\mathfrak{so}(3)}
\newcommand{\posif}{p^{\mathcal{I}}}
\newcommand{\dposif}{\dot{p}^{\mathcal{I}}}
\newcommand{\velif}{v^{\mathcal{I}}}
\newcommand{\dvelif}{\dot{v}^{\mathcal{I}}}

\newcommand{\gravbfr}{g^{\mathcal{B}}}

\newcommand{\gravifr}{g^{\mathcal{I}}}
\newcommand{\posbfr}{p^{\mathcal{B}}}
\newcommand{\velbfr}{v^{\mathcal{B}}}
\newcommand{\accelbfr}{a^{\mathcal{B}}}

\newcommand{\magifr}{m^{\mathcal{I}}}

\newcommand{\vect}{\mathrm{vec}}

\newcommand{\xbfr}{x^{\mathcal{B}}}

\newcommand{\se}{\mathrm{SE}_5(3)}
\newcommand{\am}{\alpha^{m}}
\newcommand{\ap}{\alpha^{p}}
\newcommand{\av}{\alpha^{v}}

\newcommand{\zbfr}{z^{\mathcal{B}}}

\begin{document}
\maketitle 


\begin{abstract}
This paper studies nonlinear observer design for rigid-body extended pose estimation using inertial measurements and generic exteroceptive sensing. The estimation problem is formulated as a cascade architecture that separates translational dynamics from rotational kinematics while preserving the geometric constraint of attitude evolution on $\so$. By embedding the inertial navigation model into a Linear Time-Varying (LTV) representation, we construct an observer composed of a Kalman-Bucy-type estimator for translational states and an auxiliary unconstrained attitude variable, coupled with a nonlinear geometric reconstruction filter evolving on $\so$. The cascade interconnection is analyzed within a nonlinear systems framework. We prove that uniform observability of the LTV subsystem guarantees almost global asymptotic stability of the overall nonlinear observer. For a benchmark GPS–landmark-aided configuration, explicit sufficient conditions on admissible trajectories are derived to ensure uniform observability. Simulation results illustrate the effectiveness of the proposed estimation framework.
\end{abstract}

\section{Introduction}
Accurate estimation of a rigid body's extended pose, including position, velocity, and orientation, is a critical requirement in many robotics and aerospace applications \cite{Titterton2004StrapdownTechnology}. Inertial Navigation Systems (INS), which rely on accelerometers and gyroscopes embedded within an Inertial Measurement Unit (IMU), are commonly used to perform this task. These systems integrate sensor data to estimate the vehicle's state over time. However, this approach is prone to drift due to sensor noise, biases, and unknown initial conditions, leading to inaccurate state estimates if IMU data are used alone \cite{Woodman_INS_tech_report,goel2021introduction}. To mitigate these issues, INS is often augmented with external measurements, such as those from the Global Positioning System (GPS), which provides periodic corrections to the position estimates. In GPS-denied scenarios, such as in indoor environments, alternative sensors like magnetometers, cameras, or acoustic systems are employed to obtain supplementary measurements. For example, vision-aided INS integrates IMU data with visual information from cameras to improve state estimation accuracy \cite{Marco_ECC_2020, reis2018source, Single_bearing_ECC2024, wang2020hybrid}. By fusing these complementary data sources, the system can compensate for the limitations of individual sensors, thereby enhancing overall reliability.

Despite the effectiveness of traditional Extended Kalman-type filters in fusing multiple sensor inputs, these approaches are limited by their reliance on local linearization, making them sensitive to initial estimation errors and less robust under nonlinear dynamics \cite{ReedJesse , multiple0trakingKim2022, EKF0Marina0application0Zhao}. Invariant Kalman filters \cite{barrau2017invariant,barrau2018invariant} have emerged as a reliable and generic solution, providing local asymptotic stability while overcoming some of the challenges faced by traditional methods. Recent research has focused on nonlinear deterministic observers that provide stronger stability guarantees and are well-suited for handling the nonlinearities inherent in INS applications (see, \textit{e.g.,} \cite{mahony2008nonlinear, izadi2014rigid, zlotnik2016nonlinear, bryne2017nonlinear, Berkane_Automatica_2021, Wang_TAC_2022}). While most nonlinear observer designs for INS are tailored to specific sensor configurations, limiting their applicability to a predefined set of measurements, some efforts have been made toward more general frameworks. In particular, \cite{Berkane_Automatica_2021} proposes a unified approach that accommodates generic position measurements but only in the inertial frame.

Recently, the authors proposed a geometric observer on the Lie group $\se$ in \cite{SE25_paper}, which decouples the translational and rotational error dynamics without linearization and guarantees almost global asymptotic stability. However, the admissible output set in that framework is restricted to measurements that admit a right-invariant representation on the group. In the present work, this limitation is removed by generalizing the measurement model to encompass a significantly broader class of exteroceptive observations. The proposed framework accommodates partial and heterogeneous measurements, including bearing-to-feature observations, Pitot-tube airspeed measurements, and scalar attitude information, among others. Based on this extended measurement structure, a generic cascade observer architecture for inertial navigation is developed.

A central contribution of this work is the reformulation of the inertial navigation estimation problem as a cascade interconnection between a Linear Time-Varying (LTV) subsystem and a nonlinear geometric reconstruction stage on $\so$. The LTV embedding enables the construction of a Kalman--Bucy-type observer that estimates translational states together with an auxiliary unconstrained attitude variable, while a nonlinear filter evolving on $\so$ reconstructs a physically consistent attitude. This separation preserves the geometry of the rotational dynamics while allowing the translational estimation problem to be addressed using standard linear filtering tools.

The use of LTV embeddings for observer design has appeared in specific contexts, including attitude estimation \cite{BATISTA2012_single_vector, BATISTA2012_cascaded, Batista_2012_sensor_based} and bearing-based inertial navigation \cite{Single_bearing_ECC2024}. In contrast, the proposed framework accommodates a broad class of heterogeneous exteroceptive measurements---including GPS, landmark observations, bearing measurements, Pitot-tube airspeed sensing, and scalar attitude information—within a unified cascade architecture. By analyzing uniform observability of the LTV subsystem, sufficient conditions are established to guarantee almost global asymptotic stability of the overall nonlinear observer.

\section{Notation}\label{section:Notation_Preliminaries}
We denote by $\mathds{Z}_{>0}$  the set of positive integers, by $\R$ the set of real numbers, by $\R^n$ the $n$-dimensional
Euclidean space, and by $\sphere^n$ the unit $n$-sphere embedded in
$\R^{n+1}$. We use $\|x\|$ to denote the Euclidean
norm of a vector $x\in\R^n$. The $i$-th element of a vector $x\in\R^{n}$ is denoted by $x_i$. The $n$-by-$n$ identity and zeros matrices are denoted by $I_n$ and $0_{n\times n}$, respectively. The unit vectors along the coordinate axes are $e_1$, $e_2$, and $e_3$. The Special Orthogonal group of order three is denoted
by $\so:= \{A\in\R^{3\times3}: \mathrm{det}(A) = 1; AA^{\top} =A^{\top}A=I_3\}$. The set 
$\liealg:=\{\Omega\in\R^{3\times3}:\Omega=-\Omega^{\top}\}$ denotes the Lie algebra of $\so$.  The vector of all ones is denoted by $\mathds{1}$. For $x,\ y\in\Rthree$, the map $[.]_{\times}:\Rthree\to\liealg$ is defined such that $[x]_{\times}y=x\times y$ where $\times$ is the vector
 cross-product in $\Rthree$. The Kronecker product between two matrices $A$ and $B$ is denoted by $A
 \otimes B$. The  vectorization operator $\vect:\R^{m \times n}\to\R^{mn}$ , stacks the columns of a matrix $A \in \R^{m \times n}$ into a single column vector in $\R^{mn}$. The inverse of the vectorization operator, $\vect_{m,n}^{-1}:\R^{mn}\to\R^{m\times n}$, reconstructs the matrix from its vectorized form by reshaping the $mn\times1$ vector back into an $m\times n$ matrix form. We introduce the following important orthogonal projection operator $\Pi:\sphere^2 \to \Rthree$,
defined as $
\Pi_x=I_3-xx^{\top},\quad x\in\sphere^2.$
Note that $\Pi_x$ is an orthogonal projection matrix which geometrically
projects any vector in $\Rthree$ onto the plane orthogonal to vector $x\in\sphere^2$.
In addition, one verifies that $\Pi_xy=0_{3\times1}$ if $x$ and $y$ are collinear. For simplicity and for the sake of clarity, the argument of the time-dependent signals is omitted unless otherwise required.


\section{Problem Formulation}\label{Section:Problem_formulation}
Let $\iframe$ be an inertial frame, $\bframe$ be a North-East-Down (NED) body-fixed frame attached to the center of mass of a rigid body (vehicle) and the rotation matrix $R\in\so$ be the orientation (attitude) of frame $\bframe$ with respect to $\iframe$. Consider the following 3D kinematics of a rigid body
\begin{subequations}\label{equation:dynamic_model_Inertial_frame}
\begin{align}
\label{eq:dp}
\dposif&=\velif,\\
\label{eq:dv}
\dvelif&=\gravifr+R\accelbfr,\\
\dot{R}&=R\wx,
\end{align}
\end{subequations}
where the vectors $\posif\in\R^{3}$ and $\velif\in\R^{3}$ denote the position and linear velocity of the rigid body expressed in frame $\iframe$, respectively, $\omega$ is the angular velocity of $\bframe$ with respect to $\iframe$ expressed in $\bframe$, $\gravifr\in\R^{3}$ is the gravity vector expressed in $\iframe$, and  $\accelbfr\in\R^{3}$ is the 'apparent acceleration' capturing all non-gravitational forces applied to the rigid body expressed in frame $\bframe$. 

This work focuses on the problem of position, linear velocity, and attitude estimation for INS. We assume that the vehicle is equipped with an IMU providing measurements of the angular velocity $\omega$ and apparent acceleration $\accelbfr$ (inputs). Note that the translational system \eqref{eq:dp}-\eqref{eq:dv} is a linear system with an unknown input $R\accelbfr$. Therefore, there is a coupling between the translational dynamics and the rotational dynamics through the accelerometer measurements. Most adhoc methods in practice assume that $R\accelbfr\approx -\gravifr$ to remove this coupling between the translational and rotational dynamics. However, this assumption holds only for non-accelerated vehicles, {\it i.e., } when $\dvelif\approx 0$. In this work, we instead design our estimation algorithm without relying on this assumption.

The objective of this paper is to design an almost globally asymptotically convergent observer to simultaneously estimate the inertial position $\posif$, inertial velocity $\velif$ and attitude $R$ using a set of $q \in \Zsg$ measurements, where each measurement $m_i$, $i\in\{1,\cdots,q\}$, is defined according to one of the following sensing modalities:
\begin{itemize}
\item[$(i)$] The body-frame vector measurements $m_i=R^{\top}r_i$ where  $r_i\in\R^{3}$ is constant and known,  which represents observations of a known and constant inertial vector in the body-frame (\textit{e.g.,} from a magnetometer),  \textbf{and/or}
\item[$(ii)$] The body-frame landmark measurements $m_i=R^{\top}(p_i-\posif)$ where  $p_i\in\R^{3}$ is the position of a constant and known landmark in the inertial frame, (\textit{e.g.,} from stereo vision system),  \textbf{and/or}
\item[$(iii)$] The body-frame bearing-to-landmark measurements  $m_i=\tfrac{R^\top (p_i-\posif)}{||p_i-\posif||}$, with $p_i$ constant and known, captured \textit{e.g.,} from a monocular camera, \textbf{and/or}
\item[$(iv)$] The body-frame linear velocity $m_i=R^{\top}\velif$, captured, \textit{e.g.,} from airspeed sensor or Doppler radar, \textbf{and/or}
\item[$(v)$] The body-frame scalar measurements  $m_i=d^\top_i R^\top v^{\mathcal{I}}$, captured from a Pitot tube where $d_i$ is the known direction of the Pitot probe, which represents the component of the vehicle's velocity in the direction of the probe, expressed in the \textit{body frame}, \textbf{and/or}
\item[$(vi)$] The GPS inertial position  measurements $m_i=\posif+Rb_i$, captured from a GPS located at $b_i\in\R^{3}$, a constant and known position in the body-fixed frame (lever arm), \textbf{and/or}
\item[$(vii)$] The GPS inertial velocity  measurements $m_i=\velif$, captured from a GPS, \textbf{and/or}
\item[$(viii)$] The scalar attitude measurements $m_i=\bar{e}_i^{\top}R^{\top}\underline{e}_i$, where $\bar{e}_i,\underline{e}_i\in\R^{3}$ are constant vectors in the inertial frame. Unlike standard vector measurements that require three-axis data, this formulation allows for the use of partial vector information (\textit{e.g.,} a single reliable axis of a magnetometer) or other scalar modalities, see \cite[Section~\RomanNumeralCaps{3}] {Hassan_LCSS} for more details about scalar measurements.
 \end{itemize}
\section{Main Results}\label{Section:main_result}

We present the observer architecture and the associated stability result. 
The inertial navigation dynamics and the considered measurements are embedded into a structured 
LTV model that serves as an internal estimation system. 
A continuous-time Kalman filter estimates translational states and an auxiliary attitude variable, 
which is then projected onto $\so$ through a geometric reconstruction stage. 
Almost global asymptotic stability of the overall observer is obtained under a uniform observability condition.

\subsection{Proposed Generic Observer Architecture on $\R^{15}\times \so$}
We first construct an internal estimation model in the body frame that allows a unified treatment 
of the considered measurements and leads to a structured LTV representation suitable for observer design.
Let $\posbfr=R^{\top}\posif$, $\velbfr=R^{\top}\velif$, $\gravbfr=R^{\top}\gravifr$ be the position and linear velocity of the rigid body and the gravity vector, all expressed in $\bframe$, respectively.  The rotation variable $R$ is estimated by concatenating the columns of $R^\top$:
\begin{equation}\label{equation:zB_definition}
    z^{\mathcal{B}}:=\vect(R^\top).
\end{equation}
In view of \eqref{equation:dynamic_model_Inertial_frame}, the dynamics of the considered state variables are given by the following LTV system:
\begin{subequations}\label{equation:dynamic_model_Body_frame}
 \begin{align}
\dot{p}^{\mathcal{B}}&=-\wx \posbfr+\velbfr,\\
\dot{v}^{\mathcal{B}}&=-\wx \velbfr + \gravbfr+\accelbfr,\\
\dot{z}^{\mathcal{B}}&=-(I_3\otimes\wx)  z^{\mathcal{B}}.
\end{align}
\end{subequations}
Next, we reformulate the considered measurements into the following appropriate linear output structure:
\begin{equation}\label{equation:general_output}    y_i=C_i^pR^{\top}\posif+C_i^vR^{\top}v+C_i^{L}R^{\top}C_i^R
\end{equation}
for some matrices $C^p_i$, $C^v_i$, $C^L_i$ and $C_i^R$, $i\in\{1,\cdots,q\}$.
Table~\ref{tab:compatible_measurments} shows the corresponding matrices for each case of the considered measurements. Note that the output does not necessarily coincide with the raw measurement, since preprocessing may be required to express it in the above form. Such transformations can modify the noise characteristics and the output covariance should be adjusted accordingly during the tuning process.
\begin{table*}[]
    \centering
    \begin{tabular}{c|l|l|c|c|c|c}
        & Measurement & Output& $C^p_i$ & $C^v_i$ & $C^L_i$ & $C_i^R$\\
        
       \hline
       Vector measurements & $m_i=R^{\top}r_i$ & $y_i=m_i$ & $0_{3\times3}$ & $0_{3\times3}$ & $I_3$ & $r_i$ \\
       \hline
       Landmarks  & $m_i=R^\top (p_i-\posif)$ & $y_i=m_i$ &  $-I_3$ & $0_{3\times3}$ & $I_3$ & $p_i$ \\
       \hline
       Bearing-to-landmarks  & $m_i=\frac{R^\top (p_i-\posif)}{||p_i-\posif||}$ &  $y_i=\Pi_{m_i}R^{\top}(p_i-\posif)=0$
       & -$\Pi_{m_i}$  & $0_{3\times3}$ & $\Pi_{m_i}$ & $p_i$ \\
      \hline
      Doppler Velocity Log  (DVL)& $m_i=R^{\top}\velif$ & $y_i=m_i$ & $0_{3\times3}$ & $I_3$ & $0_{3\times3}$ & $0_{3\times1}$ \\
      \hline
      
      Pitot tube & $m_i=d^{\top}_iR^{\top}\velif$ & $y_i=m_i$ & $0_{3\times3}$ & $d^{\top}_i$ & $0_{3\times3}$ & $0_{3\times1}$ \\
      \hline
      
       GPS inertial position  &  $m_i=\posif+Rb$ & $y_i=R^{\top}(m_i-\posif)=b$ & $-I_3$ & $0_{3\times3}$ & $I_3$ & $m_i$ \\ 
      \hline
      GPS inertial velocity &  $m_i=\velif$ & $y_i=R^{\top}m_i-R^{\top}\velif=0$ & $0_{3\times3}$ & $-I_3$ & $I_3$ & $m_i$\\        
        \hline
      Scalar attitude & $m_i=\bar{e}_i^{\top}R^{\top}\underline{e}_i$ & $y_i=m_i$ & $0_{3\times3}$ & $0_{3\times3}$ & $\bar{e}_i^{\top}$ & $\underline{e}_i$ \\
    \end{tabular}
    \caption{Summary of compatible exteroceptive measurements and their LTV reformulation. This table categorizes the diverse sensor modalities supported by the proposed framework (\textit{e.g.,} landmarks, GPS, Camera, Pitot tube). For each measurement type, the original nonlinear model is presented alongside its reformulated linear output and the corresponding structural matrices $C_i^p$, $C_i^v$, $C_i^L$, and $C_i^R$. These matrices map the specific sensor data into the generic linear output equation \eqref{equation:general_output}, enabling the unification of different sensors within the single LTV Kalman filter design.}
    \label{tab:compatible_measurments}
\end{table*}

Using the identity $\vect(ABC) = (C^\top \otimes A)\vect(B)$, the measurements can be written in the 
linear form
\begin{equation}\label{equation:reformulated_output_landmarks}
    y_i = C^p_i \posbfr + C^v_i\velbfr + ((C^R_i)^{\top} \otimes C^L_i )z^{\mathcal{B}}, 
\end{equation}
where $i\in\{1,\cdots,q\}$. We can now define the following \textit{artificial} output vector:
\begin{equation}\label{equation:output_vector}
    y = \begin{bmatrix}y_1\\
                       \vdots \\
                       y_q
    \end{bmatrix}.
\end{equation}
This form of the output will be useful in the design of the observer as well as in conducting the corresponding uniform observability analysis. 
Combining the dynamics \eqref{equation:dynamic_model_Body_frame} with the output vector \eqref{equation:output_vector}, we define the state vector:
\begin{equation}\label{equation:state_the_vector}
    \xbfr = \begin{bmatrix}
        \posbfr \\
        \velbfr \\
        z^{\mathcal{B}}
    \end{bmatrix}\in\R^{15},
\end{equation}  we obtain, in view of \eqref{equation:dynamic_model_Body_frame},  \eqref{equation:output_vector} and \eqref{equation:state_the_vector}, an LTV system of the form:
\begin{subequations}\label{equation:LTV_state_model}
    \begin{align}
    \dot{x}^{\mathcal{B}}&=A(t)\xbfr+B\accelbfr,\\
    y&=C(t)\xbfr,
    \end{align}
\end{subequations}
with matrices $A(t), B$ and $C(t)$ given by
\begin{align*}
A(t)&=\bar{A}\otimes I_3+S(t),\ S(t)=
-I_5\otimes\wx,\\
\bar{A}
&=\begin{bmatrix}0 &1 &0_{1\times3} \\
0&0&(\gravifr)^{\top}\\
0_{3\times1} & 0_{3\times1} &0_{3\times3}
\end{bmatrix}\in\R^{5\times 5}, \
 \\
B&=\begin{bmatrix}
    0 & 1 & 0 &0&0
\end{bmatrix}^{\top}\otimes I_3,\\
C(t)&=\begin{bmatrix}
C^p_1 & C^v_1 &  (C^R_1)^{\top} \otimes C^L_1 \\
\vdots & \vdots & \vdots \\
C^p_q & C^v_q &  (C^R_q)^{\top} \otimes C^L_q 
\end{bmatrix}.
\end{align*}

Note that both $A(t)$ and $C(t)$ are time-varying, since they depend respectively on the angular 
velocity profile $\omega(t)$ and on the time-varying measurements. We assume throughout that these 
matrices are continuous and uniformly bounded so that the LTV system 
\eqref{equation:LTV_state_model} is well posed.

We now use this embedded LTV representation as the internal model for observer design. 
In particular, we adopt a continuous-time Kalman-type estimator, which is a standard construction 
for LTV systems \cite{besanccon2007overview,Hamel2017PositionMeasurements}. The state of 
\eqref{equation:LTV_state_model} is estimated as
\begin{align}\label{equation:Riccati_observer_full_state}
\dot{\hat{x}}^{\mathcal{B}}&=A(t)\hat{x}^{\mathcal{B}}+B\accelbfr+K(t)(y-C(t)\hat{x}^{\mathcal{B}}),
\end{align}
where $\hat x^{\mathcal{B}}$ encodes the estimates of $x^{\mathcal{B}}$ and the observer gain is given by $K(t)=PC(t)^{\top}Q(t)$
where $P$ is a positive definite matrix solution of the following Riccati equation:
\begin{equation}\label{equation:Riccati_equation}
    \dot{P}=A(t)P+PA^{\top}(t)-PC^{\top}(t)Q(t)C(t)P+M(t),
\end{equation}
The initial matrix $P(0)$ is positive definite, and $Q(t)$ and $M(t)$ are uniformly positive definite matrices to be specified. In the Kalman filter context, $M(t)$ and $Q^{-1}(t)$ represent covariance matrices of the process and measurement noises, respectively. We now recall the uniform observability condition, expressed via the observability Gramian, which guarantees uniform global exponential stability of $x^{\mathcal{B}}-\hat{x}^{\mathcal{B}}=0$,  see~\cite{Hamel2017PositionMeasurements} for details.
\begin{definition}\textbf{(Uniform Observability)}\label{definition:uniform_observability} The pair $(A(t)$$,C(t))$ is uniformly  observable if there exist constants $\delta,\mu>0$ such that $\forall t\geq0$
\begin{equation*}\label{equation:conditon_of_uniform_observability}
    W(t,t+\delta):=\frac{1}{\delta}\int_t^{t+\delta}\hspace{-0.5em}\phi^{\top}(s,t)C^{\top}(s)C(s)\phi(s,t)ds\geq\mu I_n,  
\end{equation*}
where $\phi(s,t)$ is the transition matrix associated to $A(t)$ such that $\frac{d}{dt}\phi(t,s)=A(t)\phi(t,s)$ and $\phi(t,t)=I_n$.
\end{definition}
\begin{figure*}[h!]
    \centering    \includegraphics[width=0.9\textwidth]{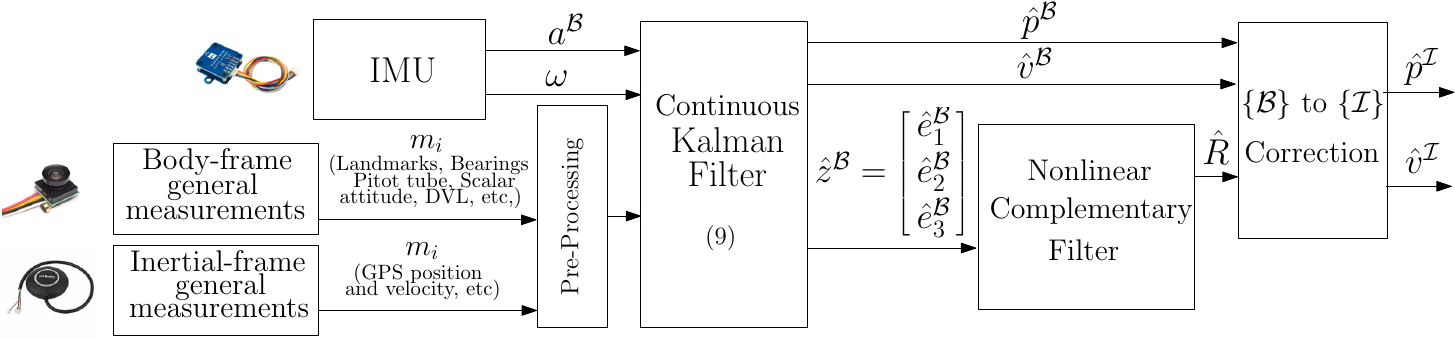}
    \caption{
    Block diagram of the proposed generic observer architecture. The framework fuses high-rate IMU inputs  with diverse exteroceptive measurements to estimate the vehicle state. A Pre-Processing stage reformulates various nonlinear sensor models into a unified LTV output structure, enabling the continuous Kalman filter to provide the body-frame estimates  $\hat{p}^{\mathcal{B}}$, $\hat{v}^{\mathcal{B}}$, and $\hat{z}^{\mathcal{B}}$. Subsequently, a nonlinear complementary filter projects the attitude estimate onto  $\so$ to recover a valid rotation matrix $\hat{R}$, which is then used to reconstruct the inertial-frame estimates $\hat{p}^{\mathcal{I}}$ and  $\hat{v}^{\mathcal{I}}$.}
    \label{figure:Illustration_of_the_proposed_method}
\end{figure*}

Let $\hat{p}^{\mathcal{B}}$, $\hat{v}^{\mathcal{B}}$, $\hat{z}^{\mathcal{B}}$, and $\hat{R}$ denote the estimates of $p^{\mathcal{B}}$, $v^{\mathcal{B}}$, $z^{\mathcal{B}}$, and $R$, respectively. The estimated body-frame directions $\hat{e}^{\mathcal{B}}_i \in \R^3$, for $i \in \{1, 2, 3\}$, are obtained by partitioning the auxiliary vector $\hat{z}^{\mathcal{B}}$ such that $\hat{z}^{\mathcal{B}} = [(\hat{e}^{\mathcal{B}}_1)^\top, (\hat{e}^{\mathcal{B}}_2)^\top, (\hat{e}^{\mathcal{B}}_3)^\top]^\top$. The overall architecture of the proposed observer is summarized in Fig.~\ref{figure:Illustration_of_the_proposed_method}. 
The LTV estimator provides the body-frame estimates $\hat{p}^{\mathcal{B}}$, $\hat{v}^{\mathcal{B}}$, and 
$\hat{z}^{\mathcal{B}}$. The latter is an unconstrained vector representation of the rotation matrix and 
is therefore not guaranteed to satisfy the orthogonality constraint. A geometric reconstruction 
stage is introduced to recover a valid attitude estimate, which is subsequently used to reconstruct the 
inertial position and velocity. In particular, we employ the following nonlinear complementary filter-like observer
\cite{mahony2008nonlinear,Wang_TAC_2022}:
\begin{equation}
  \dot{\hat{R}} = \hat{R}\,[\omega + \hat{R}^{\top}\sigma_R]_{\times},\qquad \hat R(0)\in\so,
\end{equation}
where the attitude innovation term $\sigma_R$ is given as follows:
\begin{equation}\label{equation:sigma_R}
    \sigma_R := \frac{1}{2}\hat{R}\,
\sum_{i=1}^{3} \rho_i \bigl[\hat{e}_i^{\mathcal{B}}\bigr]_{\times} (\hat{R}^{\top} e_i),
\end{equation}
with constant scalars $\rho_i>0$, $i=1,2,3$. Let the estimation errors for the augmented state $x^{\mathcal{B}}$ and its corresponding physical components be defined as:
\begin{flalign}
       \tilde{x}^{\mathcal{B}} &:= x^{\mathcal{B}} - \hat{x}^{\mathcal{B}}, \quad \tilde{p}^{\mathcal{B}} := p^{\mathcal{B}} - \hat{p}^{\mathcal{B}}, \\
            \tilde{v}^{\mathcal{B}} &:= v^{\mathcal{B}} - \hat{v}^{\mathcal{B}}, \quad 
    \tilde{z}^{\mathcal{B}} := z^{\mathcal{B}} - \hat{z}^{\mathcal{B}}, 
\end{flalign}
and the attitude estimation error be defined on $SO(3)$ as: \begin{equation}
    \tilde{R} := R\hat{R}^{\top}.
\end{equation}
The cascade interconnection between the LTV Kalman filter and the geometric reconstruction stage 
admits the following global stability property.
\begin{theorem}\label{theorem_AGAS} Consider the interconnection of system \eqref{equation:dynamic_model_Body_frame} and the nonlinear observer \eqref{equation:Riccati_observer_full_state}-\eqref{equation:sigma_R}, suppose that the pair $(A(t),C(t))$ is uniformly observable and and pick three distinct scalars $\rho_i>0$, $i\in\{1,2,3\}$. Then, the desired equilibrium point of the closed-loop system  
$(\tilde{x}^{\mathcal{B}},\tilde{R})=(0_{15\times 1},I_3)$ is almost globally asymptotically stable.
\end{theorem}
\begin{proofa}
Under the assumption of uniform observability for the pair $(A(t), C(t))$, the translational estimation error $\tilde{x}^{\mathcal{B}}$ is governed by a globally exponentially stable (GES) linear system $\dot{\tilde{x}}^{\mathcal{B}} = (A(t) + K(t)C(t))\tilde{x}^{\mathcal{B}}$. This GES subsystem acts as a vanishing perturbation to the nonlinear attitude error $\tilde{R} = R\hat{R}^{\top}$ evolving on $\so$:
$\dot{\tilde{R}} = \tilde{R} [-\psi(M\tilde{R}) - \Gamma(\hat{R})\tilde{x}^{\mathcal{B}}]_{\times}$ where, $M=\mathrm{diag}(\rho_1,\rho_2,\rho_3)$, $\psi(M\tilde{R}) = -\tfrac{1}{2}\Sigma_{i=1}^{3}\rho_i [e_i]_{\times} \tilde{R}^{\top} e_i$ and $\Gamma(\hat{R}) = \tfrac{1}{2}[0_{3\times6}, \rho_1[e_1]_{\times}\hat{R}, \rho_2[e_2]_{\times}\hat{R}, \rho_3[e_3]_{\times}\hat{R}]$. Therefore, the proof follows similar steps to \cite[Proof of Theorem 1]{SE25_paper}, see also \cite[Proof of Theorem 1] {Wang_TAC_2022}, and therefore omitted here.
\end{proofa}

Theorem~\ref{theorem_AGAS} exploits the  globally exponentially convergent property of the proposed observer and the  ISS property of the nonlinear complementary filters on $\so$ (see \cite{wang2021nonlinear})  to show that the interconnection preserves almost global asymptotic stability of the estimation errors. This is the strongest stability property achievable with smooth attitude observers.

\subsection{Structural Uniform Observability}
The stability result hinges on uniform observability of the embedded LTV pair $(A(\cdot),C(\cdot))$. 
We now show that, for a broad class of measurement configurations, this high-dimensional condition 
admits an equivalent reduced representation that greatly simplifies analysis.
\begin{lemma}\label{lemma_equivalance_of_UO_between_A_C_and_Abar_Cbar}    For all $i \in\{1,\cdots,q\}$, the configuration matrices satisfy :$$C_{i}^{p}=\alpha_{i}I_{3}, \quad C_{i}^{v}=\beta_{i}I_{3}, \quad C_{i}^{L}=\gamma_{i}I_{3}, \quad C_{i}^{R}=\theta_{i}I_{3},$$with scalars $\alpha_{i}, \beta_{i}, \gamma_{i}, \theta_{i} \in \mathbb{R}$, then the uniform observability of the high-dimensional pair $(A(t), C(t))$ is equivalent to the uniform observability of the reduced-order pair $(\bar{A}, \bar{C}(t))$, where $\bar{A}$ is defined in \eqref{equation:LTV_state_model} and \begin{equation}
    \bar{C}(\cdot)=\begin{bmatrix}
        \alpha_1 & \beta_1 & \gamma_1 & \theta_1 \\
        \vdots   & \vdots  &  \vdots  & \vdots \\
         \alpha_q & \beta_q & \gamma_q & \theta_q 
    \end{bmatrix},
\end{equation}
and $C(t)=\bar{C}(t)\otimes I_3$.
\end{lemma}
\begin{proofa}
See Appendix~\ref{appendix:proof_of_lemma_equivalance_of_UO_between_A_C_and_Abar_Cbar}.
\end{proofa}

The equivalence result provided in Lemma~\ref{lemma_equivalance_of_UO_between_A_C_and_Abar_Cbar} serves as a tool for UO analysis, reducing the complexity of observability verification in many practical scenarios since $\bar A\in\R^{5\times 5}$ is constant and has a lower dimension compared to $A(t)\in\R^{15\times 15}$. We now illustrate how this structural result can be used to 
establish uniform observability for a concrete multi-sensor configuration of practical interest.

\section{Application: Inertial Navigation Using GPS and Single Landmark Measurements}\label{section:GPS_aided_INS}
We specialize the general framework of Section~\ref{Section:main_result} to
the case where the available measurements correspond to items $(ii)$ and
$(vi)$ of Section~\ref{Section:Problem_formulation}, namely GPS inertial
position measurements $m_{gp}$ and and body-frame landmark measurements $m_{\ell}$ of a single constant and known landmark $p_{\ell}$, see
Fig.~\ref{fig:GP_landmark_scenario_illustration}. This mixed sensing setup
serves as a minimal yet nontrivial test case for applying the proposed
structural observability result. The primary objective here is to establish a sufficient condition for the global convergence of the observer described in \eqref{equation:Riccati_observer_full_state}. In view of \eqref{equation:output_vector} and Table~\ref{tab:compatible_measurments}, we derive the output vector
\begin{align}\label{equation:output_vector_GPS_INS_2}
    y &= \begin{bmatrix}
         -\posbfr + ((m_{gp})^{\top} \otimes I_3) \zbfr \\
                -\posbfr + ((p_{\ell})^{\top} \otimes I_3) \zbfr
    \end{bmatrix}.
\end{align}
Thus, we obtain the following output matrix
\begin{align}
C(t)&=\bar{C}(t)\otimes I_3,\ \bar{C}(t)=\begin{bmatrix}
    -1 & 0 & (m_{gp})^{\top}\\
    -1
 & 0 & (p_{\ell})^{\top}
\end{bmatrix}.
\end{align}


The following lemma presents a sufficient condition  to ensure the uniform observability of the pair $(A(\cdot),C(\cdot))$.
\begin{lemma}\label{Lemma:sufficient_condition_for_UO_GPS_INS}
 If there exist $\delta,\mu>0$ such that, for any $t\geq0$,
\begin{flalign}\label{equation:condition_of_lemma_GPS_INS}
  \gravifr(\gravifr)^{\top}+\frac{1}{\delta}\int_t^{t+\delta}\hspace{-1em}(p_{\ell}-\posif(\tau))(p_{\ell}-\posif(\tau))^{\top}d\tau\geq\mu I_3,
\end{flalign}

then, the pair $(A(\cdot),C(\cdot))$ is uniformly observable.
\end{lemma}
\begin{proofa}
See Appendix~\ref{appendix:proof_of_Lemma:sufficient_condition_for_UO_GPS_INS}.
\end{proofa}

The condition \eqref{equation:condition_of_lemma_GPS_INS} can be interpreted geometrically by analyzing the rank of the constituent matrices. The term $g^{\mathcal{I}}(g^{\mathcal{I}})^{\top}$ is a rank-1 matrix that spans the vertical direction (parallel to gravity). For the sum to be uniformly positive definite (\textit{i.e.,} $\ge \mu I_3$), the integral term involving the relative position must necessarily span the subspace orthogonal to gravity. Therefore, condition (20) implies that the relative position vector $(p_{\ell} - p^{\mathcal{I}})$ must be Persistently Exciting (P.E.) in the horizontal plane. In other words, the projection of the relative position onto the plane orthogonal to $g^{\mathcal{I}}$ must be two-dimensional over the integration interval $[t, t+\delta]$. Practically, this ensures observability provided the vehicle neither moves purely vertically relative to the landmark for an extended period, nor maintains a constant horizontal direction indefinitely.

The following lemma considers the case where a body-frame vector measurement of a constant and known inertial direction is available (\textit{e.g.,} from a magnetometer).

\begin{lemma}\label{lemma:with_magnetometer}
    Consider the case where, in addition to GPS and landmark measurements, a magnetometer is available providing measurements of a constant and known inertial magnetic field $m^{\mathcal{I}} \in \mathbb{R}^3$. If there exist constants $\delta, \mu > 0$ such that for all $t \ge 0$:
    \begin{flalign}
       & g^{\mathcal{I}}(g^{\mathcal{I}})^\top + m^{\mathcal{I}}(m^{\mathcal{I}})^\top + \nonumber\\&\frac{1}{\delta} \int_{t}^{t+\delta} (p_\ell - p^{\mathcal{I}}(\tau))(p_\ell - p^{\mathcal{I}}(\tau))^\top d\tau \ge \mu I_3,
    \end{flalign}
    then the pair $(A(t), C(t))$ is uniformly observable.
\end{lemma}
\begin{proofa}
    The proof follows the same derivation steps as Lemma~\ref{Lemma:sufficient_condition_for_UO_GPS_INS}.
\end{proofa}

Lemma~\ref{lemma:with_magnetometer} shows that the inclusion of a magnetometer significantly relaxes the persistence of excitation requirements on the vehicle's trajectory. Geometrically, assuming the gravity vector $g^{\mathcal{I}}$ and magnetic field $m^{\mathcal{I}}$ are non-collinear, they span a fixed two-dimensional plane, meaning the constant term in the condition already provides a rank-2 contribution. Consequently, for the Observability Gramian to be full rank, the time-varying integral term involving the relative position $(p^\ell - p^{\mathcal{I}})$ need only excite the single remaining dimension orthogonal to this plane (\textit{i.e.,} the direction aligned with $g^{\mathcal{I}} \times m^{\mathcal{I}}$). In practical terms, this implies that unlike the gravity-only case which necessitates lateral maneuvering, the magnetometer-aided configuration maintains uniform observability even during straight-line motion, provided the vehicle's relative trajectory does not remain permanently confined within the plane defined by the gravity and magnetic field vectors.


\begin{remark}
     A key challenge with the proposed observer \eqref{equation:Riccati_observer_full_state} is that the dimension of the original state space is $9$ while the dimension of the observer is $15$ potentially resulting in more restrictive conditions for uniform observability. Nevertheless, by incorporating information from the natural constraint 
$R^{\top}R=I_3$ it is possible to generate \textit{additional} virtual outputs   helping to overcome this drawback. Indeed, using the already available GPS position $m_{gp}$ and landmark measurements $m_{\ell}$, one can generate a virtual output that is compatible with \eqref{equation:general_output}. First, we have 
\begin{flalign}
    m_{\ell}\times\posbfr&=(R^{\top}(p_{\ell}-\posif))\times R^{\top}\posif\nonumber =R^{\top}[p_{\ell}-\posif]_{\times}\posif\nonumber\\
    &=R^{\top}[p_{\ell}]_{\times}\posif\nonumber
    =R^{\top}[p_{\ell}]_{\times}m_{gp}\nonumber\\
    &=((p_{\ell}\times m_{gp})^{\top}\otimes I_3)z^{\mathcal{B}}.
\end{flalign}
Hence, we can define the following virtual output:
\begin{equation}\label{equation:vitual_output_GPS_landmark}
    y=\begin{bmatrix}
    [m_{\ell}]_{\times} & 0_{3\times3} & -(p_{\ell}\times m_{gp})^{\top}\otimes I_3)    
    \end{bmatrix}x^{\mathcal{B}}.
\end{equation}
Detailed observability analysis under this additional output is omitted here. 
\end{remark}

\begin{figure}[h!]
    \centering
    \includegraphics[width=\linewidth]{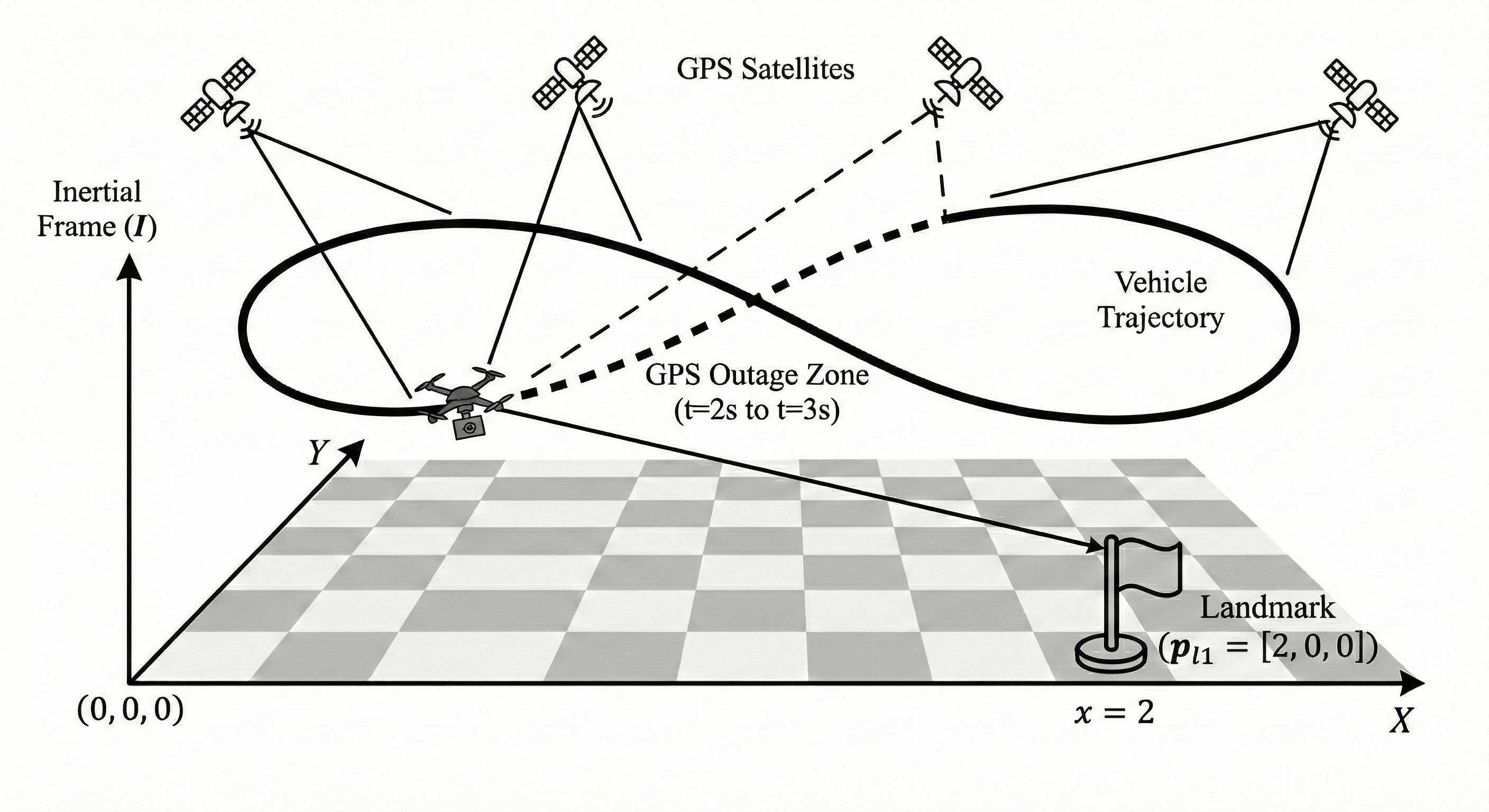}
    \caption{Illustration of GPS-Landmark INS scenario.}
    \label{fig:GP_landmark_scenario_illustration}
\end{figure}
\section{Simulation Results}\label{section:simulation}
In this section, we provide simulation results to test the performance of the observer proposed in Section~\ref{Section:main_result}.
Consider a vehicle moving in 3D space and tracking the following eight-shaped trajectory: $
    p(t)=[\cos(5t), \sin(10t)/4 , -\sqrt{3}\sin(10t)/4]^{\top}$.
The rotational motion of the vehicle is subject to the following angular velocity: $    \omega(t)=[\sin(0.3t) ,0.7\sin(0.2t+\pi) ,0.5\sin(0.1t+\pi/3)]^{\top}$. The initial values of the true pose are $\posif(0)=[
1, 0 ,0]^{\top}$, $\velif(0)=[
    -0.0125, 2.5, -4.33]^{\top}$, $R(0)=\exp([\pi e_2]_{\times}/2)$ and $\gravifr=[ 0, 0 , 9.81]^{\top}$.  

Three scenarios are analyzed: $(i)$ when only the GPS position is measured (GPS), $(ii)$ when both GPS position and landmark are measured (GPS-Landmark) and $(iii)$ when in addition to $(ii)$ virtual output \eqref{equation:vitual_output_GPS_landmark} is incorporated (GPS-Landmark-Virtual). 
The landmark is located at $p_{\ell}=[2, 0, 0]^{\top}$. 
The proposed observer gains are set to $P(0) = I_{15}$, $V = 100 I_{15}$ while the matrices $Q$ are specified in Table~\ref{matrix_Q}, and $\rho_i=100$, $i=1,2,3$.
\begin{table}[ht]
\centering
\caption{Measurement noise covariance matrices ($Q$) for the considered observer configurations.}
\label{tab:noise_covariances}
\begin{tabular}{lccc}
\toprule
\textbf{} & \textbf{GPS} & \textbf{GPS-Landmark} & \textbf{GPS-Landmark-Virtual} \\ 
\midrule
$Q$ & $0.05 I_3$ & $0.05 I_6$ & $\begin{bmatrix} 0.05 I_3 & 0.05 I_3 & 0.01 I_3 \end{bmatrix}^{\top}$ \\ 
\bottomrule
\end{tabular}
\label{matrix_Q}
\end{table}

We perform $100$ Monte Carlo runs. Initial estimation errors are randomly generated from Gaussian distributions: the initial relative attitude $\hat{R}(0)$ corresponds to an average error of $20^{\circ}$ per rotation axis with a $30^{\circ}$ standard deviation, the average initial relative position error is $\tilde{p}(0) = [-4.5, -5, 6]^\top $  with standard deviation 3 per axis, and the average initial relative velocity error is $\tilde{v}(0) = [1, -1.5, 0.5]^\top $  with standard deviation 1 per axis.

 The measurements are considered to be affected by a Gaussian noise of a noise-power of $10^{-1}$
for IMU, GPS and landmark measurements. The simulation results are presented in Fig.~\ref{figure:errors_simulation_result} and Table~\ref{tab:rmse_comparison}. The shaded areas illustrate the 5th-95th percentile of error across all Monte Carlo runs. A GPS loss event is simulated between $t=2$s and $t=3$s, indicated by the vertical dashed lines. 
\begin{itemize}
    \item During normal operation ($t < 2$s): All observers converge rapidly from initial errors as the active GPS signal serves as a position "anchor," rendering landmark measurements redundant for position accuracy and resulting in overlapping curves. The GPS-Landmark-Virtual observer (green) shows comparable or slightly improved performance during this phase due to the additional geometric constraint, while the landmark's primary utility shifts toward correcting attitude errors.
    \item During the GPS outage ($2\text{s} \le t \le 3\text{s}$): The unassisted GPS-only observer (blue) loses its anchor and diverges significantly as it relies purely on dead reckoning. In contrast, the landmark-aided observers maintain bounded errors by switching reliance to the landmark's relative position measurements to constrain drift; notably, the virtual measurement becomes unavailable without GPS, causing the Virtual (green) and Standard Landmark (red) curves to overlap.
    \item Upon GPS return ($t > 3$s): Upon the restoration of the GPS signal, the standard GPS observer re-converges, while the landmark-aided observers continue to track the state with high precision.
\end{itemize}

Overall, the inclusion of landmark measurements significantly improves robustness against GPS outages. The virtual measurement provides an additional theoretical constraint that conditions the covariance matrix effectively when GPS is available, though it reduces to the standard landmark update during GPS loss.


\begin{figure}[h!]
    \centering
    \includegraphics[width=1\linewidth]{figures/figures_simulation/monte_carlo_position.png}
\includegraphics[width=1\linewidth]{figures/figures_simulation/monte_carlo_velocity.png}
\includegraphics[width=1\linewidth]{figures/figures_simulation/monte_carlo_attitude.png}
    \caption{The estimation errors for inertial position (top), velocity (middle), and attitude (bottom) over time. A temporary GPS Loss is simulated between $t=2\,\text{s}$ and $t=3\,\text{s}$, followed by a GPS Return.}
    \label{figure:errors_simulation_result}
\end{figure}
\begin{table}[h]
\centering
\caption{Overall Root Mean Square Error (RMSE) Comparison}
\label{tab:rmse_comparison}
\begin{tabular}{lccc}
\toprule
\textbf{Configuration} & \textbf{Position} & \textbf{Velocity} & \textbf{Attitude} \\ \midrule
GPS                    & 3.3366            & 7.1169            & 0.8193            \\
GPS-Landmark           & 2.2355            & 3.8487            & 0.3510            \\
GPS-Landmark-Virtual   & 2.1023            & 3.5292            & 0.3012            \\ \bottomrule
\end{tabular}
\end{table}
\section{Conclusion}\label{section:conclusion}
This paper proposed a cascade observer architecture for inertial navigation
that combines an LTV Kalman–Bucy-type estimator with a nonlinear geometric
reconstruction filter on $\so$. By embedding a broad class of measurement
models into a unified LTV structure, the framework enables systematic
observer construction while preserving geometric consistency. Under derived
uniform observability conditions of the LTV subsystem, the cascade
interconnection was shown to achieve almost global asymptotic stability.

The GPS–single-landmark case study highlights both theoretical and practical
implications of the approach. Although GPS measurements dominate accuracy
during nominal operation, simulations demonstrate that a single landmark is
sufficient to prevent unbounded drift during temporary GPS outages. The
relative geometric constraint maintains bounded estimation errors and allows
rapid recovery once GPS returns, showing that even minimal environmental
structure can significantly enhance robustness of inertial navigation.

A structural limitation of the LTV embedding is that the observer evolves on
a higher-dimensional manifold than the original state. We showed that the
orthogonality constraint $R^\top R = I_3$ can be exploited to construct
additional virtual outputs that mitigate this drawback and improve
observability conditioning in practice. Future work will extend the framework
to biased inertial sensors and long-duration navigation scenarios.

\section*{Appendix}

\subsection{Proof of Lemma~\ref{lemma_equivalance_of_UO_between_A_C_and_Abar_Cbar}}\label{appendix:proof_of_lemma_equivalance_of_UO_between_A_C_and_Abar_Cbar}

Let us first compute the state transition matrix for \eqref{equation:LTV_state_model}. Let $\mu>0$ and  $a^{\mathcal{B}}\equiv0$. Consider the change of variable $x(t)=T(t)\xbfr(t)$ where $T(t)=I_5\otimes R(t)$, for any $t\geq0$. Then, by direct differentiation one obtains $\dot x =\bar{A}x$ which implies that $x(t)=\exp(\bar{A}(t-s))x(s)$ for any $0\leq s \leq t$. Therefore, $\xbfr(t)=T(t)^{\top}\exp(\bar{A}\otimes I_3(t-s))T(s)\xbfr(s)$ which implies that the state transition matrix is given by 
\begin{align}\label{equation:proof:transition_matrix}
\phi(t,s)&=T(t)^{\top}\exp(\bar{A}\otimes I_3(t-s))T(s)\nonumber\\
&=T(t)^{\top}(\exp(\bar{A}(t-s))\otimes I_3)T(s)\nonumber\\
&=:T(t)^{\top}\big(\bar{\phi}(t,s)\otimes I_3\big)T(s). 
\end{align}
\normalsize
Assume that the pair $(\bar{A},\bar{C}(t))$ is uniformly observable. We now show that the observability Gramian of the pair $(A(t),C(t))$ satisfies the condition of Definition~\ref{definition:uniform_observability}. The observability Gramian for the pair $(A(t),C(t))$ is written as $
    \textstyle W(t,t+\delta)=\frac{1}{\delta}\int_{t}^{t+\delta}\phi^{\top}(s,t)C^{\top}(s)C(s)\phi(s,t)ds,$ from which we obtain in view of \eqref{equation:proof:transition_matrix}, 
\begin{flalign}\label{equation:proof:grammian_matrix_with_phi_bar}
W(t,t+\delta)=T^{\top}(t)\frac{1}{\delta}\int_{t}^{t+\delta}\mathcal{O}^{\top}(s,t)\mathcal{O}(s,t)dsT(t),
\end{flalign}
\normalsize
where $\mathcal{O}(s,t)=C(s)T(s)^{\top}(\bar{\phi}(s,t)\otimes I_3)$.
On the other hand, since $C(t)=\bar{C}(t)\otimes I_3$, $T(t)=I_5\otimes R(t)$, we have $T(t)C^{\top}(t)C(t)T(t)^{\top}=(I_5\otimes R(t))(\bar{C}^{\top}(t)\otimes I_3)(\bar{C}(t)\otimes I_3)(I_5\otimes R^{\top}(t)) $. Therefore, using the Kornecker product property $(A\otimes B)(C\otimes D)=AC\otimes BD$, for any matrices $A,B,C,D$ of appropriate dimensions, and the fact that $R\in\so$, we obtain  $
    T(s)C^{\top}(s)C(s)T(s)^{\top}=\bar{C}^{\top}(s)\bar{C}(s)\otimes I_3.$
Hence, it follows that, $
   \mathcal{O}^{\top}(s,t)\mathcal{O}(s,t)=
    \big(\bar{\phi}^{\top}(s,t)\bar{C}^{\top}(s)\bar{C}(s)\bar{\phi}(s,t)\big)\otimes I_3$. 
Substituting this result back into \eqref{equation:proof:grammian_matrix_with_phi_bar}, we obtain $W(t,t+\delta)=T^{\top}(t)\big(\bar{W}(t,t+\delta)\otimes I_3\big) T(t),$ where $\bar{W}(t,t+\delta)=\textstyle\frac{1}{\delta}\int_{t}^{t+\delta}\bar{\phi}^{\top}(s,t)\bar{C}^{\top}(s)\bar{C}(s)\bar{\phi}(s,t)ds$. Since the pair $(\bar{A}, \bar{C}(t))$ is uniformly observable, there exist $\mu>0$ such that $\bar{W}(t,t+\delta)\geq \mu I_5$. Therefore, $W(t,t+\delta)\geq \mu I_{15}$, and thus, the pair $(A(t),C(t))$ is also uniformly observable. The converse direction is straightforward.


\subsection{Proof of Lemma~\ref{Lemma:sufficient_condition_for_UO_GPS_INS}}\label{appendix:proof_of_Lemma:sufficient_condition_for_UO_GPS_INS}
We start by showing that the pair $(\bar{A}, \bar{C}(t))$ is uniformly observable. Define the following matrix 
\footnotesize
\begin{align}\label{equation:O_M}
\mathcal{O}=\begin{bmatrix}
   N_0\\N_1\\N_2
        \end{bmatrix}:=\begin{bmatrix}
            \bar{C}(t)\\
N_0(t)\bar{A}+\dot{N_0}(t)
\\
N_1(t)\bar{A}+\dot{N_1}(t)
        \end{bmatrix}
\end{align}
\normalsize
Let $\mu,\delta>0$, we first show that, for any $t\geq0$
\footnotesize
\begin{flalign}\label{equation:proof_of_lemma_4_OO_block_matrix}
\frac{1}{\delta}\int_t^{t+\delta}\mathcal{O}^{\top}(\tau)\mathcal{O}(\tau)d\tau\geq\mu I_3.
\end{flalign}
\normalsize
Without loss of generality, we consider that the measurements are not noisy and thus, $m_{gp}=\posif$, $m_{\ell}=R^\top (p_{\ell}-\posif)$. We have from \eqref{equation:O_M}
\footnotesize
\begin{align}
\mathcal{O}=\begin{bmatrix}
      -1 & 0 & (m_{gp})^{\top}\\
    -1  & 0 & (p_{\ell})^{\top}\\
 0& -1 & (\velif)^{\top}\\
  0& -1 & 0 \\
  0 & 0 & (\dot{v}^{\mathcal{I}}-\gravifr)^{\top}\\
  0 & 0 & -(\gravifr)^{\top}
        \end{bmatrix}.
\end{align}
\normalsize
from which we obtain after raw operations
\footnotesize
\begin{align}
\mathcal{O}=\begin{bmatrix}
      -1 & 0 & (m_{gp})^{\top}\\
   0  & 0 & (p_{\ell})^{\top}\\
 0& -1 & (\velif)^{\top}\\
  0& -1 & 0 \\
  0 & 0 & (\dot{v}^{\mathcal{I}})^{\top}\\
  0 & 0 & -(\gravifr)^{\top}
        \end{bmatrix}.
\end{align}
\normalsize
we have
\footnotesize
\begin{equation}\label{equation:proof_of_lemma_4_OO_transpose}
    \mathcal{O}^{\top}(t)\mathcal{O}(t)=\left[
\begin{array}{c:c}
  \begin{matrix}
      1&0\\0&1
  \end{matrix} & Y_1(t) \\
  \hdashline \\  [-1em] 
  Y_1^{\top}(t) & Y_2(t) \\
\end{array}
\right],
\end{equation}
\normalsize
where $Y_1^{\top}(t)=\begin{bmatrix}
      -\posif(t) & 0
\end{bmatrix}$, $Y_2(t)=\posif(t)(\posif(t))^{\top}+(p_{\ell}-\posif(t))(p_{\ell}-\posif(t))^{\top}+\velif(t)(\velif(t))^{\top}+( \dot{v}^{\mathcal{I}}(t))(\dot{v}^{\mathcal{I}}(t))^{\top}+( \gravifr)(\gravifr)^{\top}$. Now, let $\textstyle
 W(t,t+\delta)=\frac{1}{\delta}\int_t^{t+\delta}\mathcal{O}'^{\top}(\tau)\mathcal{O}'(\tau)d\tau.$
In view of \eqref{equation:proof_of_lemma_4_OO_transpose}, one obtains
\footnotesize
\begin{flalign}
 W(t,t+\delta)=\left[
\begin{array}{c:c}
   \begin{matrix}
      1&0\\0&1
  \end{matrix} & \frac{1}{\delta}\int_t^{t+\delta}Y_1(\tau)d\tau \\
  \hdashline  \\  [-1em] 
 \frac{1}{\delta}\int_t^{t+\delta}Y_1^{\top}(\tau)d\tau & \frac{1}{\delta}\int_t^{t+\delta}Y_2(\tau)d\tau 
\end{array}
\right].\nonumber
\end{flalign}
\normalsize
The Schur complement of $W$ is given by
\footnotesize
\begin{flalign}
      \bar{W}&(t,t+\delta)= \frac{1}{\delta}\int_t^{t+\delta}Y_2(\tau)d\tau -\\ &\frac{1}{\delta^2}\int_t^{t+\delta}Y_1^{\top}(\tau)d\tau \begin{bmatrix}
      1&0\\0&1
    \end{bmatrix} \int_t^{t+\delta}Y_1(\tau)d\tau, \nonumber
\end{flalign}
\normalsize
from which we derive, using Cauchy-Schwartz inequality, that
\footnotesize
\begin{flalign}\label{equation:proof_lemma_GPS_after_shur_complement}
    \quad \quad\bar{W}(t,t+\delta)=&\frac{1}{\delta}\int_t^{t+\delta}(p_{\ell}-\posif(\tau))(p_{\ell}-\posif(\tau))^{\top}d\tau+\nonumber\\&\frac{1}{\delta}\int_t^{t+\delta}\velif(\tau)(\velif(\tau))^{\top}d\tau+\gravifr(\gravifr)^{\top}+\\&\frac{1}{\delta}\int_t^{t+\delta}\dot{v}^{\mathcal{I}}(\tau)(\dot{v}^{\mathcal{I}}(\tau)))^{\top}d\tau\nonumber  
\end{flalign}
\normalsize
Hence, \eqref{equation:proof_lemma_GPS_after_shur_complement} yields
\footnotesize
\begin{flalign}\label{equation:last_inequlity_proof_of_lemma_4}
  \quad \quad\bar{W}(t,t+\delta)\geq&\frac{1}{\delta}\int_t^{t+\delta}(p_{\ell}-\posif(\tau))(p_{\ell}-\posif(\tau))^{\top}+\gravifr(\gravifr)^{\top}  
\end{flalign}
\normalsize
In view of  the inequality in Lemma~\ref{Lemma:sufficient_condition_for_UO_GPS_INS} and \eqref{equation:last_inequlity_proof_of_lemma_4}, one obtains $\bar{W}(t,t+\delta)\geq\mu I_3$, for any $t\geq0$. It follows also that \eqref{equation:proof_of_lemma_4_OO_block_matrix} is satisfied. Moreover, since $\bar{A}$ is constant with real eigenvalues, it follows in view of \cite[Proposition 2.4]{hamel2020correction} that the pair $(\bar{A},\bar{C}(t))$ is uniformly observable. Finally, in view of Lemma~\ref{lemma_equivalance_of_UO_between_A_C_and_Abar_Cbar}, we conclude that the pair $(A(t),C(t))$ is also uniformly observable, which concludes the proof.
\bibliographystyle{unsrt}
\bibliography{main}
\end{document}